\documentstyle[12pt,aps,fleqn]{revtex}
\input{psfig}

\newcommand{\beq}{\begin{eqnarray}}
\newcommand{\eeq}{\end{eqnarray}}
\begin{document}
\begin{center}
{\bf\LARGE Diffusion of a Deformable Body in a Randomly Stirred Host Fluid}\\
\vspace{.7cm}
{\bf  Moshe Schwartz, Gady Frenkel}\\
{\em Raymond and Beverly Sackler Faculty of Exact Sciences\\
School of Physics and Astronomy\\
Tel Aviv University, Ramat Aviv, 69978, Israel}
\end{center}
\begin{abstract}
We consider a deformable body immersed in an incompressible liquid
that is randomly stirred. Sticking to physical situations in which the
body departs only slightly from its spherical shape, we calculate the
mean squared displacement and the diffusion constant of the body. We give explicitly the dependence of
the diffusion constant on the velocity correlations in the liquid and
on the size of the body. We emphasize the particular case in which the
random velocity field follows from thermal agitation.
 
\end{abstract}
%
\section{Introduction}
%
%
Systems of deformable objects immersed in a liquid are very common in
every day life. Milk and blood, for example, are such composite
systems. Milk can be viewed as an emulsion of fat globules in water
while blood is a suspension of cells ( that have some rigidity) in
water. The physical description of the set of objects present in a
given liquid involves the location of the objects, their shape and
in some cases the strains on the objects or any other fields that are
needed to describe the objects in addition to their location and
shape. The actual solution of such systems is extremely difficult
because each object interacts with itself and with the other objects
via hydrodynamic interactions. Hence, we are facing a many body
problem with the additional complication, that each object is not
described by a single degree of freedom ( its center of mass) but
actually by an infinite number of degrees of freedom, where all the deformation
degrees of freedom, corresponding to all the objects, interact. The
situation is simplified a little if the deviation of the objects from
spherical shape remains small \cite{cox,biesel}. This happens when the
agitation of the 
host liquid is not too strong and when the density of the objects is
not too high ( Close packing would cause finite
deformations, although perhaps still treatable within the small
deformation approximation). Our final goal is to obtain the response
of the composite system to a given velocity field imposed on the
liquid. The velocity field we have in mind may be fixed in time like
simple shear or randomly fluctuating in time and space. Even in the
first case the velocity field experienced by each object separately
must have a random part due to the random passage of other objects near
by. The present article is the first in a series that deals with this
general problem and it concentrates, within the small deformation
approximation, on the diffusion of the center of mass of a deformable
body in the presence of a random external velocity field, imposed on
the liquid. It is possible to consider the center of mass separately
from the deformation degrees of freedom, that will be discussed in a
future publication, because as will be shown in the following, they
are decoupled in the small deformation approximation. We study the
mean square displacement (MSD) of an object as a function of time for
a general random velocity field that has given correlations in space
and time. For long periods of time the MSD is usually linear in time, enabling
us to discuss it in terms of a diffusion constant that depends on the
size of the object, $R$, and the correlations present in the
liquid. There are, of course, cases where the MSD does not behave
linearly at long times and the method we develop here is quite capable of
dealing with those cases too. Our main concern is in objects in which
the state of lowest energy is of spherical shape. This is the case for
deformable objects dominated by surface tension \cite{navot99}. Our
results results 
concerning the motion of the center of the object will hold also for
cases where the shape of lowest energy is nearly spherical (for
example a body with bending energy \cite{helfrich76,lisy98} and
spontaneous curvature close to that of a sphere with the same volume).

The paper is organized in the following way:
In section \ref{sec:system} we define the system under consideration,
discuss its properties, and construct the basic equations.
In section \ref{sec:derivation} we derive the equation of motion for the center and the MSD in integral and differential forms.
In section \ref{sec:results} we discuss the results and demonstrate
their use for different kinds of noise realizations. The specific case
of thermal agitation is considered in section \ref{sec:thermal}. In
appendix \ref{appendix:thermal}, we construct the velocity correlations
for the case of thermal agitation and in appendix
\ref{sec:aproximation} we consider the physical conditions under which
the small deformation approximation is valid.

\section{The System} \label{sec:system}
\begin{figure}[!htp]
\centerline{\psfig{figure=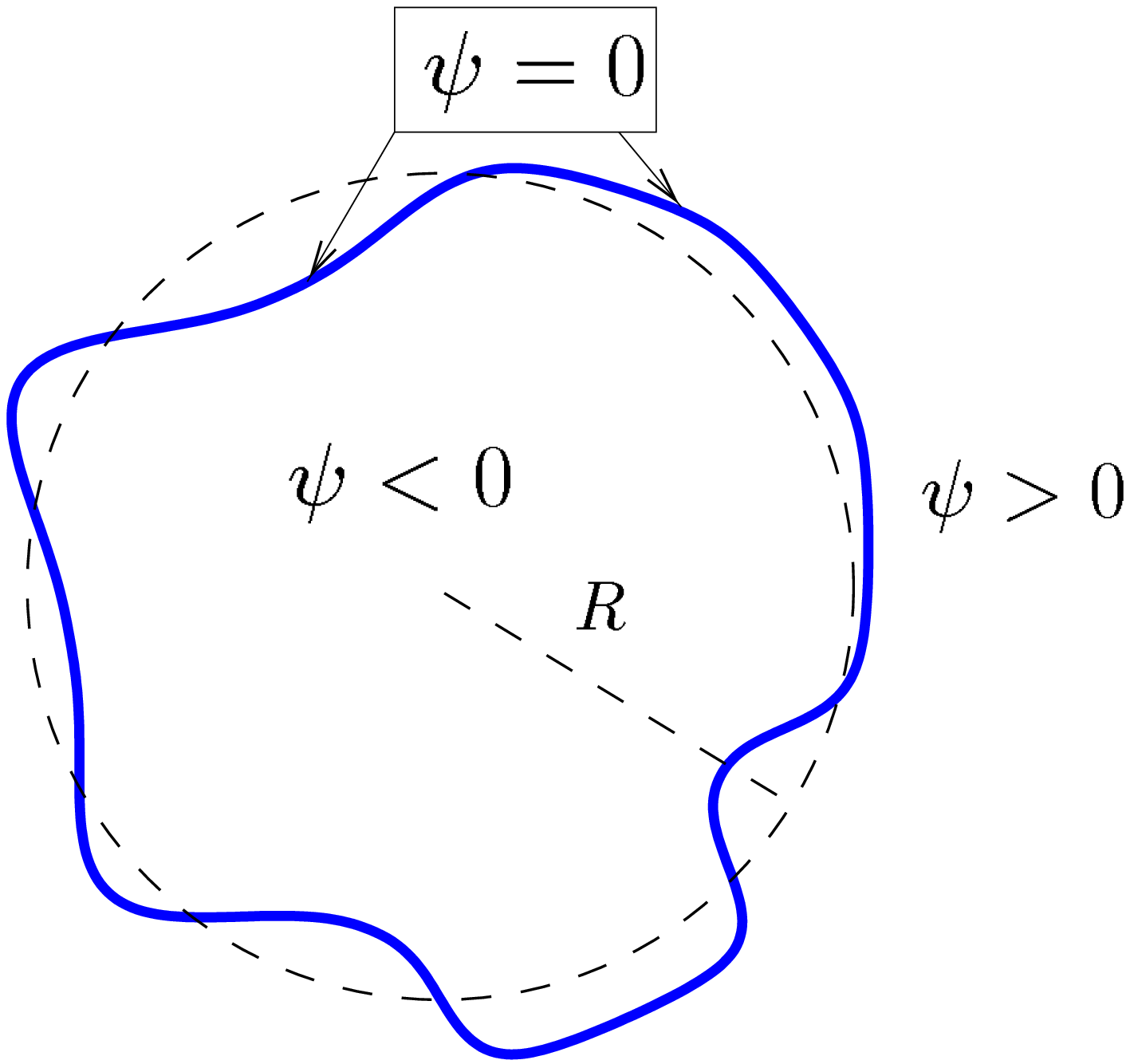,width=18cm ,height=17cm,clip=}}
%
\caption{}

\label{fig:system}
\end{figure}

%
%
Consider a single deformable body immersed in a host fluid.
The system is chosen to have the following characteristics:\\
\begin{enumerate}
\item The host fluid is incompressible,
$\vec{\nabla}\cdot\vec{v}=0$. In addition, we assume that the Reynolds
number is small so that the Stokes approximation is applicable.
\item The body is characterized by an energy that depends on its
shape. The shape of minimum energy is a sphere.
The shape of the surface of the body is described by the equation: 
$\psi(\vec{r})=0$ where $\psi(\vec{r})$ is a scalar three dimensional
field (Fig. \ref{fig:system}).( Although our derivation considers only
objects for which the shape of minimum energy is a sphere, all the
conclusions concerning the MSD carry over to cases where the shape of
minimum energy is nearly spherical).
\item A deformation of the body induces a force density on the host
fluid. As a result, the fluid's velocity is given by the sum of
$\vec{v}_{ext}$, which is caused by external sources, and
$\vec{v}_\psi$, which is induced by the body.
\item The external velocity, $\vec{v}_{ext}$, is random  
and is chosen to have zero average and known
correlations. It is convenient to define
the external velocity in terms of its spatial Fourier transform as
\beq  \label{eq:vext}
v_{ext_i}(\vec{q}\ ) \equiv \sum_j \left( \delta_{ij}-\frac{q_i
    q_j}{q^2}\right) u_j(\vec{q}),
\eeq
where $\vec{u}(\vec{q})$ is a general vector field and the subscripts
denote Cartesian components. This definition implies
only that the fluid is incompressible and in any other way is general.
Next, invariance under translations in space and time and under
rotations yields the form of the
correlations of the velocity, 
\beq 
\left\langle u_l(\vec{q},t)\right\rangle = 0 \mbox{  and }  \nonumber
\eeq
\beq   \label{eq:fcorelation}
\left\langle u_l(\vec{q},t_1) u_m(\vec{p},t_2) \right\rangle =
\delta_{lm}\delta(\vec{q}+\vec{p})\phi(q, |t_2-t_1|) ,
\eeq 
where $\delta_{lm}$ is the Kronecker delta and $\delta()$ is the Dirac
delta function and 
$\phi$ is a general function of $q$ and $|t_2-t_1|$. If the random velocity field is
characterized by a length scale and a characteristic time scale then
it is convenient to write it as $\phi(\xi q, |t_2-t_1|/\tau)$, where
$\xi$ and $\tau$ are, respectively, the 
correlation length and the memory time scale of the external velocity .

\item The surface elements of the body are carried by the
host fluid \cite{schwartz90b}, i.e. each surface point moves according
to 
\beq   \label{eq:surfaceelement}
\dot{\vec{r}} = \vec{v}_{ext}(\vec{r}) + \vec{v}_\psi(\vec{r}).
\eeq
\item We assume that the external velocity is weak enough to cause
only minor shape fluctuations of the body.\\
\end{enumerate}
%
%
%
%
%

We will be interested in the following in the Mean Squared
Displacement (MSD) of the center. Since the body is deformable the
definition of its center is not unique. For periods of time shorter than
$\tau$ the result depends on the definition of the center. It turns
out, however, that the value of the MSD at 
longer times does not depend on the specific choice, because for long
times the MSD (according to any reasonable definition) is much larger than the
size of the body. Therefore, the results for the diffusion constant
are general and do not depend on the specific definition of the center
which will be determined later.In cases where the long time dependence
of the MSD is not linear, it is still tending to infinity with time,
so that again the specific definition of the center does not matter.\\

Following the line of derivation of Schwartz and Edwards
\cite{schwartz90b,schwartz88}, equation (\ref{eq:surfaceelement}) may be
turned into a continuity equation for $\psi$,
\beq  \label{eq:continuity}
\frac{\partial \psi}{\partial t} + (\vec{v}_{ext} + \vec{v}_\psi)\cdot
\vec{\nabla}\psi = 0 .
\eeq
Consider a deformable body, carried by the host fluid in such a way
that at any instant it is nearly spherical. Its state can thus be
characterized by the position of its center, $\vec{r}_0(t)$, and a
deformation function $f(\Omega,t)$ that describes the shape by the
equation
\beq \label{eq:parameterization}
\psi(\vec{r},t) \equiv \frac{\rho}{R} + f(\Omega,t) -1 = 0,
\eeq
where $\rho \equiv |\vec{r} - \vec{r}_0|$ is the distance of the surface from the center in the direction of the
solid angle $\Omega$ and $R$ is the radius of the body when not
deformed. The deformation  function $f$ can be expanded in spherical
harmonics, $f(\Omega,t) = \sum_{l=0}^\infty \sum_{m=-l}^l
f_{l,m}(t)Y_{l,m}(\Omega)$. The center of the shape, $\vec{r}_0(t)$, is
defined as that point around which $f_{1m}(t)=0$.\\ 
Equations (\ref{eq:continuity}) and (\ref{eq:parameterization}) lead
to a linear equation for each $f_{l,m}$,
\beq  \label{eq:flmdot}
\frac{\partial f_{lm}}{\partial t} + \lambda_l f_{lm}
+\frac{1}{R}[\hat{\rho}\cdot(\vec{v}_{ext} - \dot{\vec{r}}_0)]_{lm}=0  ,
\eeq
where $\hat{\rho}$ is a unit vector directed outwards from the center
in the direction of $\Omega$, and
\beq \label{eq:Qlm}
[\hat{\rho}\cdot(\vec{v}_{ext} - \dot{\vec{r}}_0)]_{lm} =
\int d\Omega \left\{
    \hat{\rho}\cdot\left[\vec{v}_{ext}(\vec{r}_0 + R(1-f)\hat{\rho}) -
      \dot{\vec{r}}_0\right]Y^*_{l,m}(\Omega)\right\} .
\eeq
The eigenvalues $\lambda_l$'s characterize the decay of a slightly
deformed sphere into a sphere in the absence of the external
velocity. Different physical systems are characterized by different
sets of $\lambda_l$'s. However, it is
obvious that the decay must depend only on $l$ because of the spherical
symmetry. Examples of systems for which different sets of $\lambda_l$
have been calculated include: a droplet with a surface tension and equal
viscosities inside and outside \cite{schwartz88} and a droplet with
surface tension for a viscosity much higher inside the droplet than
outside \cite{gang}. Other systems for which the following results are
applicable to, in the small deformations approximation, include a droplet with a
bending energy \cite{safran}, a droplet with a
bending energy and in-plane dissipation \cite{foltin} and a droplet
with both surface tension and bending energy \cite{komura}.
Our following discussion is therefore general and not limited to one
specific system.
%
%
%
%
\section{Derivation of the MSD} \label{sec:derivation}
%
%
%
%
Equation (\ref{eq:flmdot}) implies that in order that $f_{1,m}$
stays zero for all times we must have as an equation determining the
location of the center
\beq   \label{eq:motion1}
[\hat{\rho}\cdot(\vec{v}_{ext} - \dot{\vec{r}}_0)]_{1m} = 0 \ \ , \
m=-1,0,1 .
\eeq

For $l \neq 1$ it is clear that $\dot{\vec{r}}_0$ can be dropped from
the last term on the left hand side of
Eq. (\ref{eq:flmdot}). Therefore $f(\Omega,t)$ is linear in
$\vec{v}_{ext}$ (for long enough times the initial deformations have
already decayed). Consequently we can always drop, for small enough
$\vec{v}_{ext}$,  $f$ in the argument of
$\vec{v}_{ext}$ on the right hand side of Eq. (\ref{eq:Qlm}) (The
physical conditions for which this approximation is valid are
discussed in appendix \ref{sec:aproximation}). This
results in decoupling of the deformation degrees of freedom from that
of the center of the sphere. The equation for the motion of the center
can thus be given, using linear combinations of $Y_{1,m}$, in vector form as
\beq
\int d\Omega \ \hat{\rho} \ 
\left(\hat{\rho}\cdot\dot{\vec{r}}_0 \right) =
\int d\Omega \ \hat{\rho} \ 
\left(\hat{\rho}\cdot\vec{v}_{ext}(\vec{r}_0 + R\hat{\rho})\right).
\eeq
%
We integrate the left
hand side of the above and express the external velocity in terms of
its Fourier transform on the right hand side to obtain
\beq
\frac{4\pi}{3}\dot{\vec{r}}_0 = \int d\Omega \int d^3q \ \hat{\rho}
\left(\hat{\rho}\cdot\vec{v}_{ext}(\vec{q},t)\right)e^{-i\vec{q}\cdot(\vec{r}_0+R\hat{\rho})}.
\eeq
We use the partial waves expansion \cite{bohm,landau}:
\begin{eqnarray}
e^{-i\vec{q}\cdot(R\hat{\rho})} = \sum_{l=0}^{\infty} \sum_{m=-l}^{l}
(-i)^l 4\pi j_l(qR)Y_{lm}^*(\Omega_q)Y_{lm}(\Omega) ,
\end{eqnarray}
where $\Omega$ and $\Omega_q$ are the solid angles in the directions of
$\hat{\rho}$ and $\vec{q}$ respectively and $j_l$ is the spherical Bessel
function of order $l$. We
integrate over $\Omega$ and obtain
\begin{eqnarray} \label{eq:rDotAndA}
\dot{\vec{r}}_0 &=& 3 \int  d\vec{q} 
   e^{-i\vec{q}\cdot\vec{r}_0} \left( \frac{1}{3}j_0(qR) + 
   j_2(qR){\bf A}\right)\vec{v}_{ext}(\vec{q},t) .
\end{eqnarray}
The matrix ${\bf
  A}(\vec{q})$ is given by
\beq  \label{eq:matrixAij}
A_{ij}=-\frac{2}{3}\delta_{ij} +
\left(\delta_{ij} -\frac{q_i q_j}{q^2} \right).
\eeq
It may seem that ${\bf A}$ on the right hand side of
Eq. (\ref{eq:rDotAndA}) mixes directions. However, the bracketed
term in Eq. (\ref{eq:matrixAij}) is just a projection operator
on the transverse direction. The external velocity is incompressible
and hence already transverse. Consequently, this term acts as a unity
operator, $\delta_{ij}$, and  Eq. (\ref{eq:rDotAndA}) leads to 
\begin{eqnarray} \label{eq:velocity}
\dot{\vec{r}}_0 &=&  \int  d\vec{q} 
   e^{-i\vec{q}\cdot\vec{r}_0} \left( j_0(qR) + 
   j_2(qR)\right)\vec{v}_{ext}(\vec{q},t) .
\end{eqnarray}
Equation (\ref{eq:velocity}) is the explicit equation of motion for the
center of the body. In the limit $ R \rightarrow 0$ the 
approximation, $\dot{\vec{r}}_0 = \vec{v}_{ext}(\vec{r}_0,t)$ is obtained.
Note that this equation is general and describes the motion of the
center for any given (small enough) external velocity field.\\
%
%
%
%

%
Next, we calculate the MSD, $\langle \left( \Delta\vec{r}_0
  \right)^2 \rangle$, as a function of the elapsed time, $t$. 
Consider a specific realization of the
external velocity field.

The Displacement of the center is given by the trivial equation 
\beq
\Delta \vec{r}_0(t) = \int_0^t \dot{\vec{r}}_0(t')dt'.
\eeq
Hence, the MSD is given by
\beq
\left\langle \left( \Delta\vec{r}_0(t) \right)^2 \right\rangle =
 \int_0^t dt_1 \int_0^t dt_2 \ \left\langle
  \dot{\vec{r}}_0(t_1) \cdot \dot{\vec{r}}_0(t_2)
\right\rangle.
\eeq
\\
The correlations of the external velocity given in $q$ space are
obtained from Eq.(\ref{eq:vext}) and (\ref{eq:fcorelation}) .
%
%
%
%
Assuming the decomposition \cite{gady,brus88}
\beq
\left\langle v_{ext_i}(\vec{q}_1,t_1) v_{ext_j}(\vec{q}_2,t_2)
e^{-i\vec{q_1}\cdot\vec{r}_0(t_1)}e^{-i\vec{q_2}\cdot\vec{r}_0(t_2)}
\right.  \left.\right\rangle = \nonumber \\
\left\langle v_{ext_i}(\vec{q}_1,t_1) \right.\left.v_{ext_j}(\vec{q}_2,t_2) \right\rangle \left\langle
e^{-i\vec{q_1}\cdot\vec{r}_0(t_1)}e^{-i\vec{q_2}\cdot\vec{r}_0(t_2)}
\right\rangle
\eeq
and in addition that the distribution of $\Delta\vec{r}_0(t)$ is Gaussian, i.e.
\beq
\left\langle e^{-i\vec{q}\cdot \Delta\vec{r}_0(t)}\right\rangle = 
 e^{-\frac{q^2}{6}\left\langle (\Delta r_0(t))^2\right\rangle} ,
\eeq
we obtain
\beq
\left\langle \left( \Delta\vec{r}_0(t) \right)^2 \right\rangle =& & 
\int_0^t dt_1 \int_0^t dt_2 \int d\vec{q}
  e^{-\frac{q^2}{6}\left\langle\left(\vec{r}_0(t_1)-\vec{r}_0(t_2)\right)^2\right\rangle} \nonumber\\
& &
\left( j_0(qR) + j_2(qR) \right)^2
  \sum_i\left( 1 - \frac{q_i^2}{q^2} \right)\phi(q, |t_2-t_1|).
\eeq
The only term that depends on angel is $1-q_i^2/q^2$.
Performing the angular integration $
\int d\Omega_q \left(1- q_i^2/q^2\right) = 8\pi/3$, and summing up the
three terms we obtain, denoting the MSD by $F(t)$, 
%
%
%
%
%
\beq \label{eq:MSDintegral}
F(t)=
16\pi \int_0^{t} dt' \int_0^\infty q^2dq
 e^{-\frac{q^2}{6}
F(t')}
 \phi (q,t')\left(j_0(qR)+j_2(qR)\right)^2(t-t').
\eeq
%
%
Eq. (\ref{eq:MSDintegral}) can be turned also to a differential
equation. 
Differentiating Eq. (\ref{eq:MSDintegral}) twice we obtain
\beq \label{eq:MSDdifferential}
\ddot{F}(t) = 16 \pi \int_{0}^{\infty} q^2dq
 e^{-\frac{q^2}{6}F(t)}
 \phi (q,t)\left(j_0(qR)+j_2(qR)\right)^2.
\eeq
The initial conditions are
\beq
F(0) = 0 
\eeq
\ \ \ \ \ \  and
\beq \label{eq:condition2}
\dot{F}(0) = 0.
\eeq
The latter condition is valid in cases where the correlation function,
$\phi(q,t)$,
is finite at $t = 0$. The only exception is the case of white noise, 
where one must carefully check the 
result of the first differentiation  and
determine $\dot{F}(0)$. (Actually (\ref{eq:condition2}) is always
correct, because any noise that is of physical origin must be
correlated in time. The widely used white noise is just a very useful
idealization of the real situation, that will
result in $\dot{F}(0)=0$ and $\dot{F}(\delta)$ having a value that is
not small for rather small $\delta$'s). The advantage of the differential form is that
its numerical solution can be easily obtained by advancing $F(t)$ in
time. Note that equation (\ref{eq:MSDdifferential}) above is not
restricted to cases that can be described in terms of a diffusion constant.
%
%
%
%
\section{Properties of the MSD} \label{sec:results}
%
%
%
%
The random velocity field may be caused by thermal agitation which is
an equilibrium phenomenon or by non-equilibrium process such as mechanical
stirring.
While Eq. (\ref{eq:MSDdifferential}) can supply, by numerical solution,
the MSD for any velocity correlation, there are families of velocity
correlations in which at least part of the solution of
(\ref{eq:MSDdifferential}) or (\ref{eq:MSDintegral}) can be obtained
analytically, rendering the process of solving for the MSD much easier. The
simplest case is where the correlations are white in time, namely
$\phi(q,t)=\tilde{\phi}(q)\delta(t)$. In those cases $F(t)$ is linear
at all times, $F(t)\equiv 3Dt$, where $D$ is the diffusion constant
and Eq. (\ref{eq:MSDintegral}) that is an equation for the function
$F(t)$ is replaced by an explicit expression for the diffusion
constant
\beq
D=\frac{8\pi}{3}\int_0^\infty q^2\ dq \ \tilde{\phi}(q)(j_0(qR) +
j_2(qR))^2 .
\eeq
A family of correlations that is a simple extension of the above, where
it is quite easy to see what is happening, is defined by
$\phi(q,t)=\tilde{\phi}(q)\tilde{G}(\frac{t}{\tau})$, were $\tilde{G}$
is a function that decays when its argument becomes of order 1. It is
clear from Eq. (\ref{eq:condition2}) that for short times the MSD must
behave as $t^2$ while for long times it must be linear in $t$, since
for $t>>\tau$ the time dependence cannot be distinguished from white
noise. The function $\tilde{\phi}(q)$, will naturally have a cut-off
factor $g(q\xi)$, where $\xi$ is the correlation
length. Clearly, the correlation length cannot be expected to be
smaller than the distance between the particles of which the fluid is
composed and not larger than the size of the system.

The MSD depends, of course, on the ratio $\gamma = \frac{R}{\xi}$. Generally
speaking, as $\gamma$ increases the slope of the MSD and 
particularly the diffusion constant
decreases. This is due to the fact that as $\gamma$ 
increases, different regions of the surface become less correlated and
move in different directions. In the limit  $\gamma \rightarrow
\infty$, the movement of the center ceases and $F(t)$
is always zero. In the limit  $\gamma \rightarrow 0$, the bracketed
Bessel term in Eq. (\ref{eq:MSDintegral}) and (\ref{eq:MSDdifferential}) can be replaced by unity (since $\lim_{qR \rightarrow 0}
\left( j_0(qR) +j_2(qR) \right) =1$). A close inspection of the
derivation reveals that this limit produces the same MSD equation as
the equation for the approximation $\dot{\vec{r}}_0
=\vec{v}_{ext}(\vec{r}_0)$. I.e. the latter approximation is accurate
for an infinite correlation length, or point particles.\\

In the following we will consider the dependence of the diffusion
constant on the size of the object $R$.
Consider a correlation such as $\phi (q,t) = C\ \delta(t)\
(q \xi)^\alpha g(q\xi)$, where $g$ is a cutoff function and $g(0)>0$. 
Note that as discussed above the results that will be obtained here
for the diffusion constant hold true also for a finite correlation time.
We insert the above correlation function into equation
(\ref{eq:MSDintegral}), then substitute $qR$ with $u$, and
obtain
\beq
D = \frac{8\pi C \xi^\alpha}{3 R^{3+\alpha}}\int_0^\infty du \ u^{2+\alpha}
 g(\frac{\xi}{R}u) [j_0(u)+j_2(u)]^2 .
\eeq
In the limit $R/\xi \rightarrow\infty$ we distinguish between two
cases: $\alpha<1$ and $\alpha>1$. Since the large $u$ dependence of
$j_0(u) + j_2(u)$ is proportional to $cos(u)/u^2$ we find that
\beq  \label{eq:Dinfinity}
D \propto \left\{ \begin{array}{ll}
\frac{C \xi^\alpha}{R^{3+\alpha}} & $for $ \alpha<1\\
\frac{C \xi}{R^4} & $for $\alpha>1.\\
\end{array} \right.
\eeq
In the opposite limit $R/\xi \rightarrow 0$ we find that regardless of
$\alpha$
\beq \label{eq:Dzero}
D \propto \frac{C}{\xi^3}.
\eeq
(Note here that we have written the power law dependence of $\phi(q)$
as $q^\alpha \xi^\alpha$ but having other dimensional constants in the
model may make $C$ depend on $\xi$, so that Eqs. (\ref{eq:Dinfinity}) and
(\ref{eq:Dzero}) may be considered only as equations that yield the
dependence of $D$ on the radius $R$).\\

The dependence of the diffusion constant on the decay time scale can
be also deduced when the correlation function is separable (i.e. the
second family). Using simple dimensional analysis,
Eq. (\ref{eq:MSDintegral}) leads to the conclusion that the diffusion
constant is linear in $\tau$ ( in addition to the possible dependence
of $C$ on $\tau$): $D \sim C \tau$.\\

There is another class of velocity correlations that is not separable
but allows the calculation of the long time behavior of the MSD. This
class is defined by a scaling form of the velocity correlations,
\beq
\phi(q,t) = C q^{-\alpha}f\left( \Gamma q t^\beta \right),
\eeq
where $\Gamma$ and $C$ are dimensional constants, $f$ is a function
with a finite decay length and $\alpha<3$. 
The solution is obtained by assuming that $F(t)=A t^\nu$. The
integrand in Eq. (\ref{eq:MSDdifferential}) has in it two functions,
each cutting the integral off at different value of $q$  that is a
function of t. The dominant cut-off at large times $t$, is the one
that cuts the integrand off at smaller $q$'s. What remains is just a
scaling argument that leads from Eq. (\ref{eq:MSDdifferential}) to the
following result:
\beq \label{eq:nonseperable1}
\nu = \left\{ \begin{array}{ll}
\tilde{\nu} & $if $ \tilde{\nu}>1\\
1  & $if $\tilde{\nu} \leq 1\\
\end{array} \right. ,
\eeq
where
\beq  \label{eq:nonseperable2}
\tilde{\nu} = \left\{ \begin{array}{ll}
4/(5-\alpha) & $if $ \tilde{\nu}>2\beta\\
2 + (\alpha -3)\beta  & $if $\tilde{\nu} \leq 2\beta\\
\end{array} \right. .
\eeq
Note that Eq. (\ref{eq:nonseperable1}) results from the fact that even
if $\ddot{F}(t)=0$ for large $t$ the leading behavior of $F(t)$ is
still linear. Note also that in Eq. (\ref{eq:nonseperable2}) the two
options have to be evaluated first in order to check which of the
conditions applies. An explicit equation for the prefactor $A$ can
also be easily obtained. We solved Eq. (\ref{eq:MSDdifferential})
numerically with $\alpha=5/3$ and $\beta=3/5$. The long time
dependence of $F(t)$ is depicted in Fig. (\ref{fig:nonseperable1}). We
see that the long time dependence is given by $F(t) \propto t^{1.2}$,
which is exactly the result predicted by
Eq. (\ref{eq:nonseperable2}). (Note that in this case $\tilde{\nu}=2\beta$).
\section{Thermal Agitation} \label{sec:thermal}

Of particular interest is the case where the fluctuations in the
velocity field are due to thermal agitation. We describe the effect of
temperature by a scalar potential $\varphi$ and a vector potential
$\vec{A}$, that fluctuate, have zero average and local correlations in
space and time. Both give rise to a force density field
\beq
\vec{F} = -\vec{\nabla}\varphi + \vec{\nabla}\times\vec{A},
\eeq
that generates in its turn the fluctuating velocity field in the
liquid. Since the velocity field is divergence-less, the scalar
potential affects only pressure. Hence, the Fourier transform of the
velocity field is given within the Stokes approximation by
\beq \label{eq:stokesq}
\vec{v}(\vec{q},t) = \frac{i\vec{q}\times\vec{A}(\vec{q},t)}{\eta q^2},
\eeq
where $\eta$ is the viscosity of the liquid. Since the correlations of
$\vec{A}(\vec{r},t)$ are local in space and time, it follows that
$\tilde{\phi}(q)$, defined by Eq. (\ref{eq:fcorelation}), is given by
\beq \label{eq:22}
\tilde{\phi}(q) = \frac{c}{q^2},
\eeq
where $c$ is a dimensional constant. Dimensional analysis reveals that
$c$ must be proportional to  
$K_B T / \eta$ (with a dimensionless proportionality
constant). A detailed calculation yields a proportionality constant
equal to $(2\pi)^{-3}$ (see appendix \ref{appendix:thermal}). The final conclusion is
that for $R$ larger than the inter-particle distance in the liquid \cite{schwartz91},
\beq
D = \frac{K_B T}{5\pi \eta R}.
\eeq
Note that this result, for a liquid membrane that has liquid inside
as well as outside, is different from the stokes result for a hard
sphere. (We may expect Eq. (\ref{eq:22}) to hold only for $q<1/\xi$ where
$\xi$ is the inter-particle distance in the liquid but since $R$ is
expected to be very large compared to $\xi$, we are always in the
situation described by Eq. (\ref{eq:Dinfinity}) with $\alpha=-2$).
The latter result is similar to the result
for a polymer subjected to thermal fluctuations \cite{doi} (with a
different pre-factor).
%
%
\appendix
\section{Velocity Correlations for Thermal Agitation} \label{appendix:thermal}
We wish to determine the exact form of the velocity correlation function for
the case of thermal agitation.
Consider a system in which the random velocity field results from
thermal agitation. The transversal part of the linearized
Navier-Stokes equation reads:
\beq
\frac{\partial v_{q}^i}{\partial t} = -\nu q^2 v_{q}^i +\frac{F^i}{\rho
  m},
\eeq
where $\vec{v}_q$ is the Fourier Transform (FT) of the velocity field,
$\nu$ is the kinematic viscosity, $\vec{F}$ is the FT of the force
density in the liquid, $\rho$ is the number density of the particles and $m$
is the mass of a liquid particle.\\
We solve this equation under the condition that the force density is
white noise in time and obtain,
\beq \label{eq:appendeq2}
\left\langle v^i_q(t_1) v^j_p(t_2) \right\rangle = e^{-\nu q^2
  |t_2-t_1|} \left\langle v^i_q v^j_p \right\rangle_{eq},
\eeq
where the last average on the right hand side is an equal time average.
It is clear from the above that the velocities at different times are
correlated, as opposed to white noise. It is possible however to
consider effective white noise correlations by integrating the right
hand side of (\ref{eq:appendeq2}) over time and replacing then the
decay function $exp(-\nu q^2 |t_2-t_1|)$ by some
$a(q)\delta(t_2-t_1)$, that will produce the same integral. This
  yields for the effective white noise velocity field,
\beq
\left\langle u_i(\vec{q},t_1) u_j(\vec{p},t_2) \right\rangle =
\frac{2}{\nu q^2} \delta(t_2-t_1)\left\langle v^i_q v^j_p \right\rangle_{eq},
\eeq
where we denote the effective velocity field by $\vec{u}$ and the real
one by $\vec{v}$. $\left\langle v^i_q v^j_p \right\rangle$ must be
proportional to $\delta(\vec{p}+\vec{q})$ because of invariance to
translations and to $[\delta_{ij} - \frac{q_i
  q_j}{q^2}]$ because of incompressibility. Comparing with
Eq. (\ref{eq:22}), we conclude that there is no additional dependence
on $q$. \\
%
%
%
%
%
Therefore,
\beq
\left\langle v^i_q v^j_p \right\rangle = c' \delta(t_2 -
t_1)[\delta{ij} - \frac{q_i q_j}{q^2}].
\eeq
Now, we wish to relate the velocity to temperature.
Considering that our continuous liquid is actually made up of $N$
discrete particles each having a mass $m$, we know that the total
kinematic energy of the liquid is $3N\frac{K_B T}{2}$ and as a result
we find
\beq
\left\langle v^2(\vec{r}) \right\rangle = \frac{3 K_B T}{m}.
\eeq
Expressing $\vec{v}$ in terms of its FT and integrating while keeping
in mind that the number of degrees of freedom should be conserved
and equal to $3N$ yields
\beq
c'=\frac{K_B T}{2m\rho (2\pi)^3}
\eeq
from which we can see that
\beq
c=\frac{K_B T}{\eta (2\pi)^3},
\eeq
and the full correlation function reads
\beq
\left\langle u_i(\vec{q},t_1) u_j(\vec{p},t_2) \right\rangle =
\frac{K_B T}{\eta (2\pi)^3}
\delta(\vec{p}+\vec{q})\delta(t_2-t_1)[\delta_{ij} -\frac{q_i
  q_j}{q^2}]\frac{1}{q^2}.
\eeq
%
%
%
%
\section{Validity of the Small Deformation Approximation} \label{sec:aproximation}
%
%
The validity of equation (\ref{eq:MSDintegral}) is limited by the
approximation of replacing 
$\vec{v}_{ext}(\vec{r}_0+ R(1+f)\hat{\rho})$ by
$\vec{v}_{ext}(\vec{r}_0+ R\hat{\rho})$ that has been discussed in
section \ref{sec:derivation}.  
The approximation implies that we can, in the limit of small deformations,
replace the velocity at the surface with the velocity on the
undeformed sphere. To check the approximation,
we expand equation (\ref{eq:motion1}) to the first nontrivial order in
$f$,
\begin{eqnarray}
\int d\Omega Y_{1,m}^*(\Omega) \hat{\rho} \cdot [\vec{v}_{ext}(\vec{r}_0 +
R\hat{\rho}) + Rf(\Omega,t)\left(\hat{\rho} \cdot \vec{\nabla}\right) \vec{v}_{ext}(\vec{r}_0 +
R\hat{\rho}) - \dot{\vec{r}}_0] =0 . \ \ \ 
\end{eqnarray}
The approximation is justified if the first order term is
negligible with respect to the zeroth order term.
A careful inspection reveals that this condition holds if 
\beq \label{eq:condition1}
R f(\Omega,t) << \xi ,
\eeq
for any spatial angle $\Omega$ at any instant of time.
The following argument is somewhat more intuitive. The deformation of
the body is of the size $R f$, and the external velocity changes at
length-scales that are comparable with the correlation length $\xi$. If
the deformation is smaller than the correlation length (Fig.
\ref{fig:approx}-A)
the external velocity does not change on the length-scale of the
deformation, and the approximation is valid. On the 
other hand, if the correlation length is shorter than the deformation
length-scale (Fig. \ref{fig:approx}-B) the external velocities on the sphere and on the body are
uncorrelated, and the approximation is unjustified.\\
We turn to evaluate $f$.
When $R<<\xi$ it is clear that $f$ can be made small by having
$V_{ext}$ small enough so that indeed $Rf<<\xi$. The
more interesting case is  $R/\xi >1$. 
The deformation $f$ is determined by Eq. (\ref{eq:flmdot})
\beq \label{append:flmeq}
\dot{f}_{l,m} + \lambda_l f_{l,m} + Q_{l,m} = 0,
\eeq
where $Q_{l,m} \equiv \frac{1}{R}[\hat{\rho}\cdot \vec{v}_{ext}]_{lm}$. 
%
Clearly,
\beq
|Q_{lm}| < \frac{4\pi}{R} \langle |Y_{lm}| \rangle v_{ext},
\eeq
where $v_{ext}$ is the typical magnitude of the external velocity. The average of the spherical harmonic is bound and of order one
and therefore can be dropped off.
$\lambda_l f_{l,m}$ is comparable with $Q_{l,m}$
(eq. \ref{append:flmeq}), therefore, in order that $Rf<<\xi$ we must
have $v_{ext} < \lambda_l \xi$. A condition that
must be true for all values of $l$ and especially for the smallest
$\lambda_l$ denoted $\lambda_{l\ min}$. Therefore,
\beq
v_{ext} < \lambda_{l\ min} \xi
\eeq
In most cases, however, we can find a stronger condition for the
validity of the small deformation approximation. We expect $Q_{l,m}$ to decline as the squared root of the number of
independent surface elements, i.e. as $\xi/R$, so that $\lambda_l f_{l,m} \sim
\frac{v_{ext}}{R} \frac{\xi}{R}$. Therefore the condition,
eq.(\ref{eq:condition1}), implies that $v_{ext} < \lambda_l R$. 
Therefore,
\beq
v_{ext} <\lambda_{l\ min} R.
\eeq
Both conditions can be easily maintained in a viscous fluid.
The above conditions are general and depend on the specific system via
$\lambda_l$.
For example for a droplet
with a surface tension energy and equal viscosities inside and
outside, the minimal eigenvalue is $\lambda_2 = \frac{16}{35}
\frac{\lambda}{\eta R}$ \cite{schwartz88} (where $\lambda$ is the surface tension constant 
and $\eta$ is the viscosity) and the condition is   
$v_{ext}<\frac{16}{35}\frac{\lambda}{\eta}$, while for a
viscosity much larger inside $\lambda_{l\ min} = \lambda_2 =
\frac{1}{2}\frac{\lambda}{\eta R}$ \cite{gang} and $v_{ext}
<\frac{1}{2}\frac{\lambda}{\eta}$ where $\eta$ is the viscosity inside
the droplet.
\newpage

{\bf Figure captions}\\

{\bf Fig. 1}\\
  The deformable body is described by a three dimensional
  scalar field $\psi(\vec{r})$. The interior is the region where
  $\psi<0$, the exterior is the region where $\psi>0$ and the outer
  surface of the body is the locus of the points obeying
  $\psi(\vec{r})=0$.\\

{\bf Fig. 2}\\
  The MSD for a fluid with a memory time scale. $F_0 \equiv \frac{C
  \tau^2}{\xi^3}$.\\

{\bf Fig. 3}\\
  The Diffusion constant for typical separable random
  velocity correlations. $D_0 \equiv \frac{C \tau}{\xi^3}$.\\

{\bf Fig. 4}\\
  The MSD for typical non-separable random
  velocity correlations with the scaling form, $\phi(q,t)= C
  q^{-\alpha} exp\left(-\Gamma q t^\beta\right)$, 
  where we choose $\alpha=\frac{5}{3}$ and $\beta =\frac{3}{5}$. The
  MSD depends  on the two non-dimensional variables:
  $\left(\frac{\Gamma}{R}\right)^{\frac{1}{\beta}}t$ and $\mu = C
  R^{\alpha-5+\frac{2}{\beta}} \Gamma^{-\frac{2}{\beta}}$. The
  MSD scales for long times as
  $t^{\frac{6}{5}}$ in agreement with our scaling argument.\\

{\bf Fig. 5}\\
  The validity condition for the approximation. Fig. A: The
  deformation is negligible in respect to the velocity correlation
  length. Therefore the approximation holds. Fig. B: The deformation length
  is longer than the correlation length. The velocities on the surfaces
  of the sphere and droplet are uncorrelated.\\

\newpage

\newpage

\begin{figure}[!htp]
\centerline{\psfig{figure=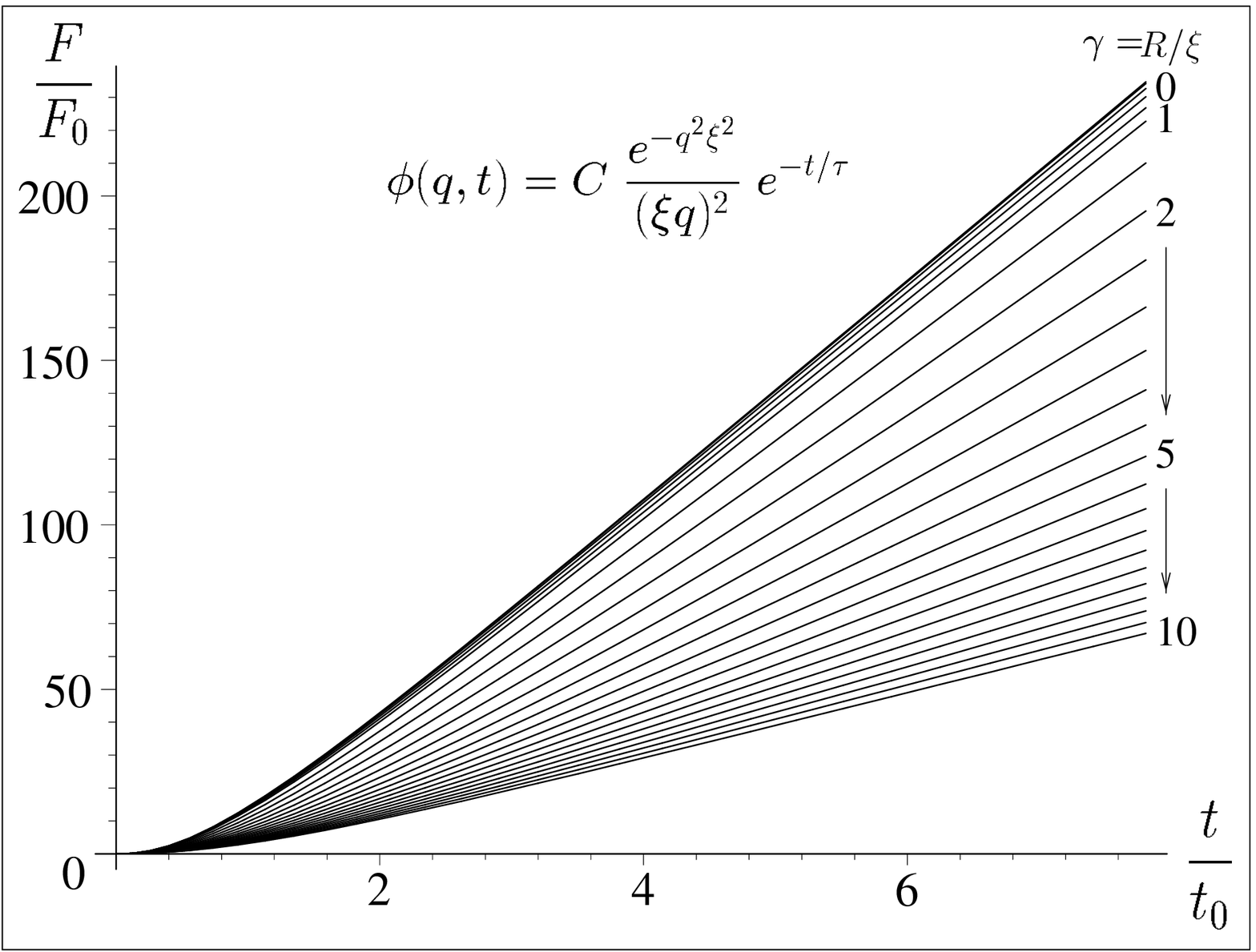,width=18cm ,height=14cm,clip=}}
\caption{}

\label{fig:expq2gausian}
\end{figure}
\newpage

\begin{figure} 
\centerline{\psfig{figure=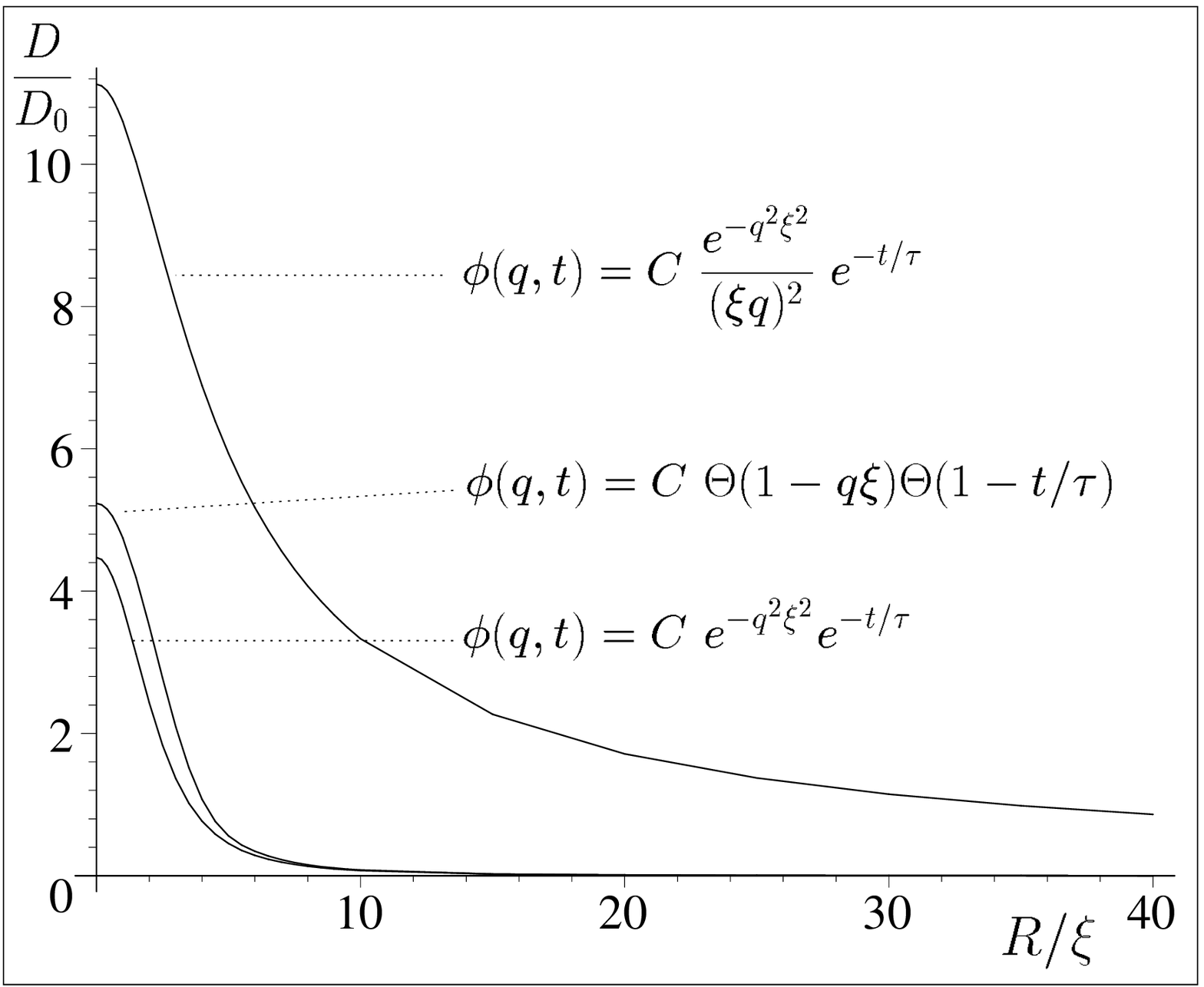,width=18cm ,height=14cm,clip=}}
\caption{}

\label{fig:Rdependence}
\end{figure} 
\newpage

\begin{figure} 
\centerline{\psfig{figure=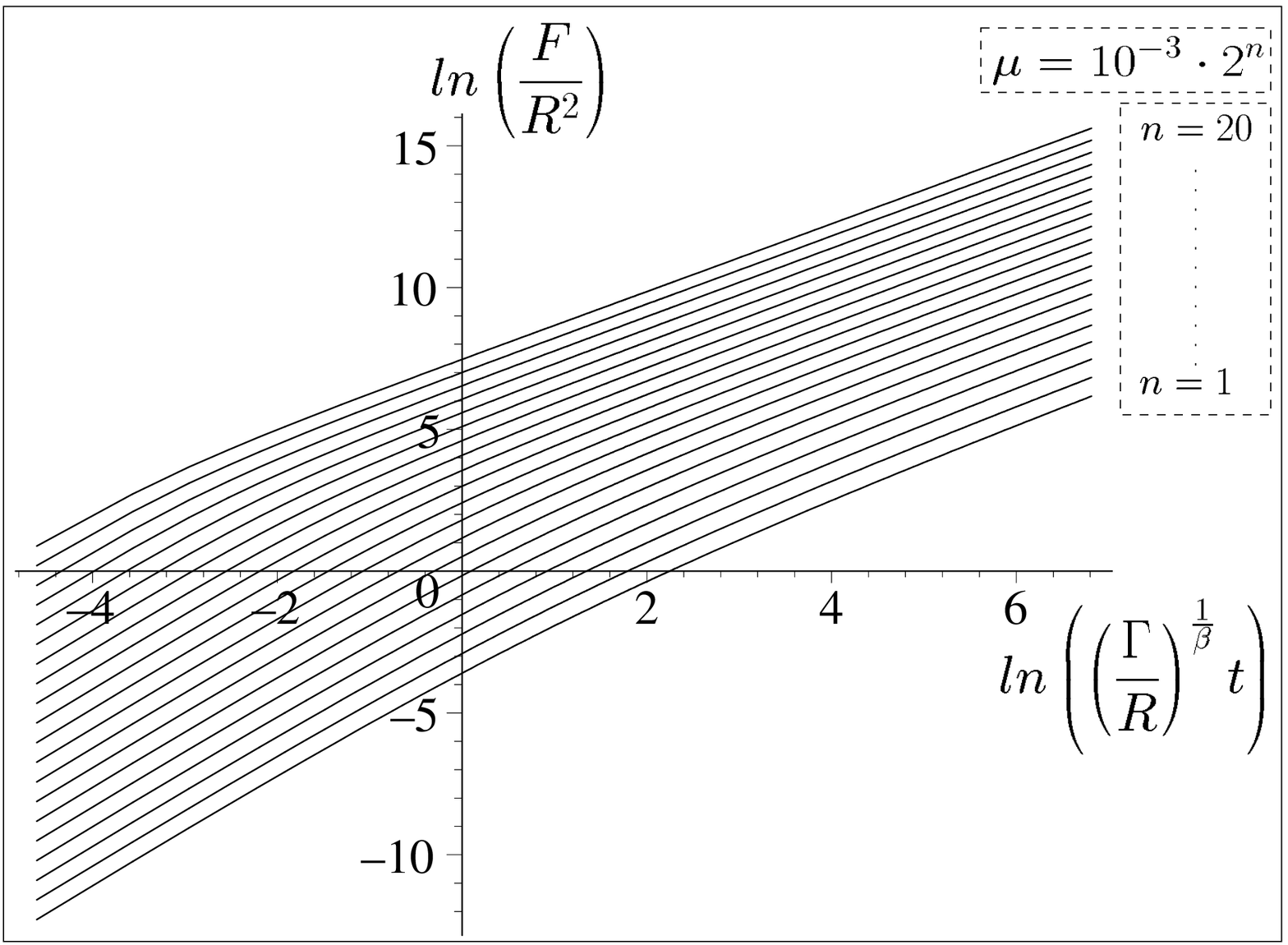,width=18cm ,height=14cm,clip=}}
\caption{}

\label{fig:nonseperable1}
\end{figure} 
\newpage

\begin{figure} 
\centerline{\psfig{figure=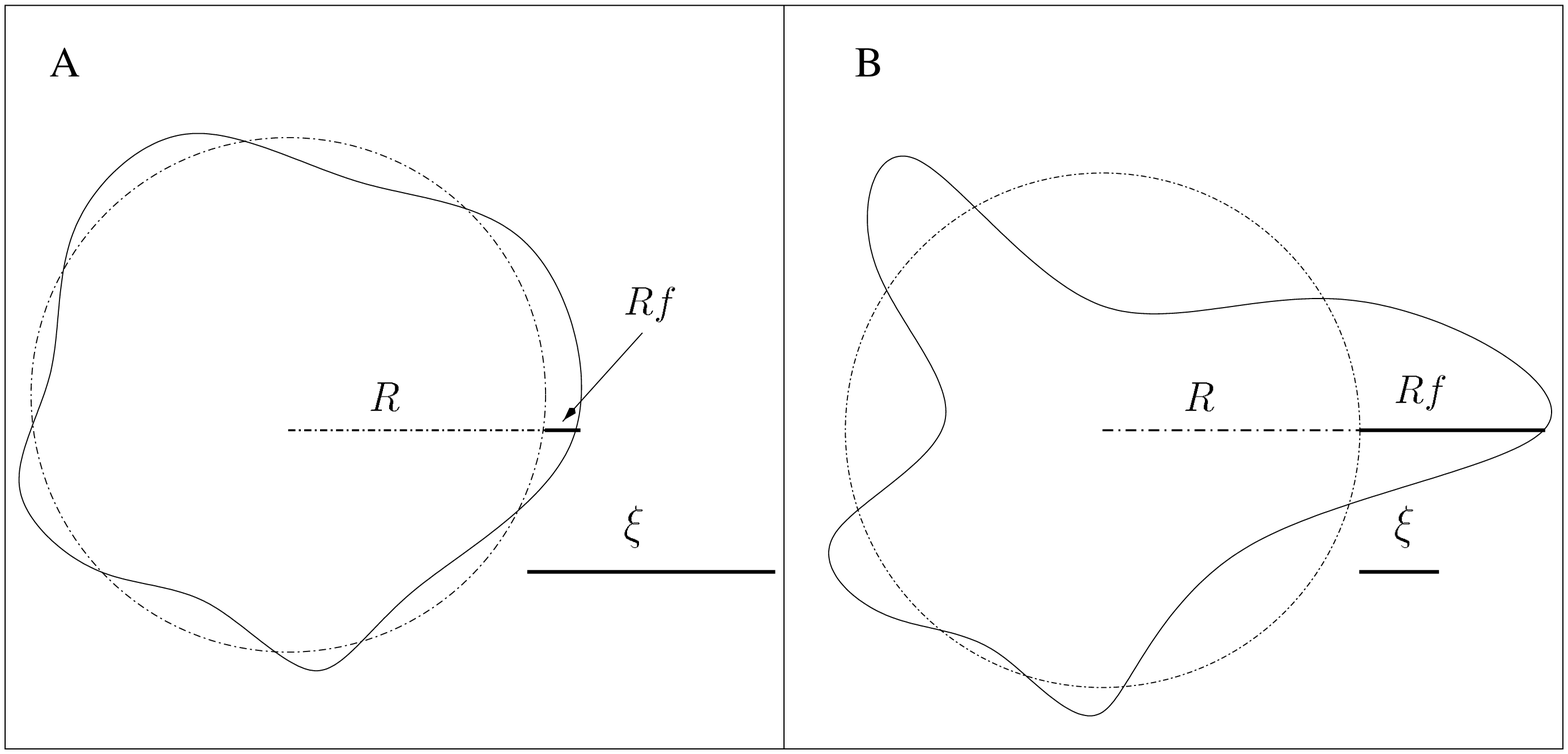,width=15cm ,height=7cm,clip=}}
\caption{}

\label{fig:approx}
\end{figure}
\end{document}